\documentstyle[twocolumn,prl,aps,epsf]{revtex}

\draft

\begin{document}

\twocolumn[\hsize\textwidth\columnwidth\hsize\csname
@twocolumnfalse\endcsname

\title{Dielectric functions and collective excitations in $MgB_{2}$}

\author{V. P. Zhukov$^1$, V. M. Silkin$^1$, E. V.Chulkov$^{1,2}$, 
 and P. M. Echenique $^{1,2}$ }

\address
{$^1$Donostia International Physics Center (DIPC), 
20018 San Sebasti{\'a}n/Donostia, Basque Country, Spain\\
$^2$Departamento de F{\'\i}sica de Materiales and Centro Mixto CSIC-UPV/EHU, 
Facultad de
Ciencias Qu{\'\i}micas, Universidad del Pais Vasco/Euskal Herriko
Unibertsitatea, Apdo. $1072, 20018$ San Sebasti\'an/Donostia, Basque Country,
Spain}

\date{\today}

\maketitle

\begin{abstract}
The frequency- and momentum-dependent dielectric function 
$\epsilon{(\bf q,\omega)}$
as well as the energy loss function Im[-$\epsilon^{-1}{(\bf q,\omega)}$\protect{]} are
calculated for intermetallic superconductor $MgB_2$ by using two {\it ab initio}
methods: the plane-wave pseudopotential method and the tight-binding version 
of the LMTO method. We find two plasmon modes dispersing at energies $\sim 2$-$8$ eV and
$\sim 18$-$22$ eV. The high energy plasmon results from a free electron like plasmon 
mode while the low energy collective excitation has its origin in a peculiar 
character of the band structure. Both plasmon modes demonstrate clearly 
anisotropic behaviour of both the peak position and the peak width. In particular,
the low energy collective excitation has practically zero width in the direction 
perpendicular to boron layers and broadens in
other directions. 
\end{abstract}

]

After the discovery of superconductivity in the $MgB_2$ compound with  
the transition temperature ${T_c}\sim39K$ \cite{naga2001} much effort has been devoted 
to understanding the mechanism of the superconductivity 
\cite{bud'ko,takahashi,kortus,an,dolgov,hirsch,voelker,osborn,guinea,liu} 
as well as to studying different 
electronic and atomic characteristics of this compound. Among these characteristics 
are the superconducting gap\cite{takahashi,kara,gorshunov}, the crystal structure and 
its influence on $T_c$ \cite{naga2001,bud'ko,sluts,jorg,saito}, band structure
\cite{kortus,an,antropov,satta,medvedeva} and the Fermi surface \cite{kortus},
lattice vibrations \cite{dolgov,osborn,taku,bohnen} as well as thermodynamic 
and transport
properties \cite{finn}. Recently Voelker {\em et al} \cite{voelker} explored 
collective excitations very near the Fermi level by using a simple band 
structure model and found the plasmon
acoustic mode at very small momenta ($q\simeq$ 0.01 $a.u.^{-1}$) and low energies
($\omega\simeq$ 0.01 eV). Here we study collective excitations in $MgB_2$ for  
different momenta and energies, namely, for $q\geq$0.1 $a.u.^{-1}$ and
$\omega \geq$ 2 eV. 
We report first-principle calculations for the real,  
$\epsilon_1{(\bf q,\omega)}$, and imaginary, $\epsilon_2{(\bf q,\omega)}$, part 
of the dielectric function as well as the energy loss function 
Im[-$\epsilon^{-1}{(\bf q,\omega)}$]. As a result of the calculation two plasmon 
modes and 
a few features arising from interband transitions are obtained.

$MgB_2$ crystallizes in the so-called $AlB_2$ structure in which $B$ atoms form
graphitelike honeycomb layers that alternate with hexagonal layers of $Mg$ atoms.
The magnesium atoms are located at the center of hexagons formed by borons
and donate their electrons to the boron planes.
Similar to graphite $MgB_2$ exhibits a strong anisotropy in the $B$-$B$ lengths:
the distance between the boron planes is significantly longer than in-plane
$B$-$B$ distance. We use this resemblance between graphite and $MgB_2$ in order 
to clear up the origin of  
plasmon peaks in $MgB_2$ by 
comparing the calculated energy loss function features with those
obtained from EELS measurements for graphite and single-wall carbon
nanotubes (SWCN) \cite{kuzuo,pichler}. To shed more light on the problem
we also evaluate the dielectric functions and the energy loss spectra for the
$MgB_2$ crystal structure with the $Mg$ atoms removed (this hypothetical crystal 
structure is designated as $B_2$). 

Information on
the energy lost by electrons in their interactions with metals is carried by
the dynamical structure factor $S(\bf q,\omega)$ which 
is related by
the fluctuation-dissipation theorem to the energy loss function 
Im[-$\epsilon^{-1}_{00}(\bf q,\omega)$]. 
To calculate the inverse dielectric function we invoke the random phase approximation
(RPA) where $\epsilon^{-1}{(\bf q,\omega)}$ is defined as (in symbolic form) 
$\epsilon^{-1} = 1+\upsilon_c{(1-\chi^0\upsilon_c)}^{-1}\chi^0$, where 
$\upsilon_c$ is the bare Coulomb potential and $\chi^0$ is the density response
function of the non-interacting electron system. The dielectric function
is related to $\chi^0$ as $\epsilon = 1-\upsilon_c\chi^0$. The energy loss function
may be obtained by inverting the first matrix 
element of $\epsilon$ that leads to neglecting short-range exchange and correlation 
effects or directly from $\epsilon^{-1}$ when these effects are included.
We have computed the energy loss function by using both of these approaches and 
found that the inclusion of the local field effects  
leads to negligible changes of both the width and energy of the plasmon peaks.  
In the calculation
of the density response matrix $\chi^{0}_{\bf GG^{'}}{(\bf q,\omega)}$ 
we have used two different first-principle methods: the plane wave 
pseudopotential method \cite{psp} and the tight-binding version of the LMTO 
method \cite{anders1984,arya1998}.

In Figs. 1a and 1b we show the evaluated band structure of $MgB_2$ and $B_2$
along the symmetry directions. In general, these band structures 
are quite similar. The distinctions between them in the vicinity of the Fermi level 
($E_F$) are due to the lower position of $E_F$ relative to the $\sigma$ band in the
$\Gamma A$ direction for $B_2$. The states of this band, which are of $p_{x,y}$
symmetry, are degenerate in $\Gamma A$ and their charge density is located in 
$B$ layer showing a clear 2D character. This character leads to weaker
interactions between the $B$ layers and to smaller dispersion of the $\sigma$
band along $\Gamma A$ in $B_2$. The $p_z$ band which is occupied at $\Gamma$ in
$MgB_2$ becomes unoccupied at $\Gamma$ in $B_2$.
In Figs. 2a and 2b we present the momentum dependence of the energy loss function 
in the $\Gamma A$, $\Gamma K$ and $\Gamma M$ directions. 
In $MgB_2$ we have found 
two plasmon modes. The higher collective excitation mode 
originates from the free electron like excitation mode with 
energy $\omega_{p1}$=19.1 eV 
that corresponds to the electron density parameter $r_s$=1.82 a.u. of $MgB_2$.
This free electron like mode 
is transformed into two separated submodes $\omega^{xy}_{p1}$ and 
$\omega^{z}_{p1}$ in a real crystal. 
One of them is very isotropic in the hexagonal plane (the $\Gamma K$ and
$\Gamma M$ directions) and disperses linearly up for momenta $q\geq 0.2$ a.u., 
while another one, in the $\Gamma A$ direction, has smaller energy and  
is nearly constant. Because of limitations of 
the calculation methods and 
a large width of the energy loss peaks at small momenta we could not 
determine accurately the plasmon peak position in this region. Therefore 
we define the plasmon energy at the $\overline{\Gamma}$ point by 
extrapolation of the computed plasmon dispersions at $q{\geq}$ 0.2 a.u.
This extrapolation results in  
$\omega_{p1}^{xy}$=19.4 eV and $\omega_{p1}^{z}$=18.8 eV in good agreement
with the free electron gas value $\omega_{p1}$. 
The width, $\bigtriangleup_{p1}$, 
of both these energy loss peaks decreases with the increasing momentum,
and the $\bigtriangleup_{p1}^{z}$ width deacreases faster than 
$\bigtriangleup_{p1}^{xy}$.

A different behavior is shown by
the low-energy loss function peak which disperses
linearly up in both the $\Gamma A$ direction and in the hexagonal plane. 
In the $\Gamma A$ direction, at $q_{\perp}\sim 0.1$ $a.u.^{-1}$, the peak 
is very narrow, 
$\bigtriangleup_{p2}^{z}\sim 0.01$ eV, 
that can be seen in the 
very small value of $\epsilon_2$ in the energy interval 
around the peak position where $\epsilon_1$=0 (Figs. 3a and 3b). 
In particular, for $q_{\perp}$=0.12 $a.u.^{-1}$
this interval is between 1 and 5 eV (Fig. 3a) and the energy loss peak
is located at 2.9 eV. On changing the momentum to the A point 
this interval 
becomes more narrow and moves to higher energies (Fig. 3c). At small momenta 
the first peak of Im[-$\epsilon_{00}{(\bf q,\omega)}$] placed in the energy
interval 0-1 eV is determined by intraband transitions within 
the two $\sigma$ bands
in the $x,y$ plane around the $\Gamma A$ direction, while the second peak
located at 5.4 eV (Fig. 3a) is determined by the interband 
$\sigma - \pi$ transitions
in the $KM$, $AH$, and $AL$ directions. So, the low energy plasmon excitation 
corresponds to electron excitations in the $\sigma$ bands and one can define
it as the $\sigma$ plasmon. 
In the hexagonal plane, in the $\Gamma K$ direction, the low energy EELS 
peak broadens (Fig. 4a) and disperses linearly up on 
going from the $\Gamma$ point to K. In the $\Gamma M$ direction the plasmon
peak disperses similar to that in the $\Gamma K$ one showing a nearly ideal 
isotropy
in the hexagonal plane. However, it becomes smaller and wider on going from
$\Gamma$ to $M$ and vanishes finally at ${q}\simeq 0.8{\mid{\Gamma M}}\mid$. 
Comparing the plasmon 
energies obtained from the LMTO and pseudopotential calculations one 
can find only a small difference of $\sim 0.1$ eV between them. 
For instance, at $q$=0.2 $a.u.^{-1}$ the LMTO 
$\omega^{z}_{p2}$ is slightly smaller than the pseudopotential one and 
$\it vice$ $\it versa$ for larger momenta. This slight difference 
results in different energy loss peak positions at $\Gamma$:
the extrapolation of both plasmon energies $\omega^{xy}_{p2}$ and
$\omega^{z}_{p2}$ calculated for $q{\geq}$ 0.1 $a.u.^{-1}$ to the $\Gamma$ point 
gives 
$\omega^{z}_{p2}$=1.8 eV, $\omega^{xy}_{p2}$=2.0 eV (LMTO)
and $\omega^{z}_{p2}$=2.2 eV, $\omega^{xy}_{xy}$=2.4 (pseudopotential).
We estimate the accuracy of these values to be better than 0.2 eV.  

Besides two plasmon modes we have obtained four small features
in Im[-$\epsilon^{-1}_{00}{(\bf q,\omega)}$] that correspond 
to interband excitations. It is difficult to find out what transitions
are responsible for 
these features, nevertheless we show them in Fig. 2a. One of them arises at
$q\simeq$0.1$a.u.^{-1}$ at an energy of $\approx$2 eV in the hexagonal plane, 
another one occurs at
$q\geq$ 0.4 $a.u.^{-1}$ in the $\Gamma M$ direction at an energy of $\simeq$ 4 eV 
and the other two small features arise in the $\Gamma A$ direction for 
$q$=0.1-0.25 $a.u.^{-1}$ at 
energies of 10 eV and 13 eV, respectively.

In Fig. 2b we show the momentum dependence of 
the energy loss function calculated for the hypothetical crystal structure $B_2$.
In general, the energy loss function in $B_2$ shows relatively similar 
features to those
in $MgB_2$, though there are some important distinctions. In particular, 
all collective 
excitations including two plasmon modes 
manifest a smaller dispersion in the $\Gamma A$ direction. This effect is a direct 
consequence of a weaker interactions between adjacent layers of boron in $B_2$
compared to $MgB_2$. Another distinction is that all features in the energy loss
function in $B_2$ are much clearer than those in $MgB_2$ (Figs. 3a-3c and 4a-4c). 
One exception is the high
energy plasmon mode. The third distinction is that $B_2$ has more features in 
Im[-$\epsilon^{-1}_{00}(\bf q,\omega)$] than does $MgB_2$. 
The low energy plasmon mode extrapolation to the $\Gamma$
point gives $\omega^{z}_{p2}$=4.1 eV which is $\sim$ 2 eV higher than that in 
$MgB_2$. 
This shift in energy is due to the higher energy position of 
the second maximum
of Im[$\epsilon_{00}$] (Fig. 3a). 
While the position of the first peak of Im$\epsilon$ in $B_2$ nearly coincides
with that in $MgB_2$ the second peak is moved by 2 eV to higher energies. 
Via the Hilbert transform (Kramers-Kronig relation) it also moves the node of 
$\epsilon_1$ to higher energy. Despite some quantitative distinctions between the 
energy loss functions in $MgB_2$ and $B_2$ one can conclude that mostly 
the features of the excitation 
spectrum of $MgB_2$ can be derived, with the relevant corrections, from those of
the hypothetical crystal $B_2$. 

The two plasmon modes similar to those obtained in $MgB_2$ were also observed 
in EELS experiments
for graphite and SWCN\cite{kuzuo,pichler} which are even
more anisotropic than $MgB_2$. The upper plasmon mode which has larger energy in 
graphite and SWCN than in $MgB_2$ (near the $\Gamma$ point $\omega^{xy}_{p1}\simeq$21 eV
(SWCN) and $\omega^{xy}_{p1}\simeq$26 eV (graphite)) \cite{pichler} also results
from excitations of all valence electrons. The lower plasmon mode $\omega^{xy}_{p2}$
shows a linear dependence on momentum, like that in $MgB_2$, with energies  
$\omega^{xy}_{p2}\simeq$5 eV (SWCN) and $\omega^{xy}_{p2}\simeq$6.5.eV (graphite)
\cite{pichler} at $\Gamma$. But in contrast to SWCN and graphite where 
$\omega^{xy}_{p2}$ represents the collective excitation of the $\pi$-electron system
\cite{kuzuo,pichler,ahuja} in $MgB_2$ this mode is a result of the collective
excitation of the $\sigma$-electron system. The different origin of the low-energy
plasmon mode in $MgB_2/B_2$ and graphite/SWCN can be qualitatively understood from
the Fermi energy ($E_F$) position. In graphite the Fermi level is pinned by 
$\pi$- electrons in the $KH$ direction. In $MgB_2$ and $B_2$ with a smaller
number of electrons per atom the Fermi level is pinned by $\sigma$-electrons. 
So, low-energy excitations in $MgB_2$ and in $B_2$ are expected to be derived 
from the $\sigma$-band electrons.

In conclusion, we have performed the first-principle calculations of the dielectric
functions $\epsilon_1{(\bf q,\omega)}$ and $\epsilon_2{(\bf q,\omega)}$ as well as
the energy loss function Im[-$\epsilon^{-1}(\bf q,\omega)$]. The calculations reveal
the two plasmon modes in $MgB_2$ and $B_2$ and a few interband collective
excitations. The low energy plasmon mode corresponding to the excitations of
electrons in the $\sigma$ bands shows a very anisotropic behavior
of the peak width.
The energy loss spectrum of $MgB_2$ can be derived, with the relevant corrections, 
from that of the hypothetical
crystal structure $B_2$.

We thank R.H.Ritchie and A.Bergara for helpful discussions. This work was partially supported by 
the Basque Country University, Basque Hezkuntza Saila,
and Iberdrola S.A.  

After the submission of this paper first-principles calculations
of collective excitations in $MgB_2$ have also been presented by Wei Ku {\it et al.}
\cite{wei}.

\begin{figure}
\caption{Calculated energy band structure of a) $MgB_2$ and b) $B_2$
along the symmetry directions. $\sigma$ and $\pi$ denote the boron 
bands of $p_{x,y}$ and $p_{z}$ character, respectively.}
\end{figure}

\begin{figure}
\caption{(color)
Dispersion of the plasmon modes (filled circles) and of the features 
(triangles) for a) $MgB_2$ and b) $B_2$ along the symmetry directions:
$\Gamma A$ (green), $\Gamma K$ (red), and $\Gamma M$ (blue).}
\end{figure}

\begin{figure}
\caption{(color) Real (dashed lines) and imaginary (solid lines) parts
of the dielectric function $\epsilon_{\bf GG}{(\bf q, \omega)}$ along 
the $\Gamma A$ direction for $\bf G$=0 and $\bf q$=(0,0,$\alpha$)(2$\pi$/c):
a) $\alpha$=1/8, b) $\alpha$=1/4, c) $\alpha$=1/2 (A point). 
The energy loss function is shown 
in the insert. The red (blue) lines represent $MgB_2$ ($B_2$).} 
\end{figure}

\begin{figure}
\caption{(color) Real (dashed lines) and imaginary (solid lines) parts
of the dielectric function $\epsilon_{\bf GG}{(\bf q, \omega)}$ along 
the $\Gamma K$ direction for $\bf G$=0 and $\bf q$=($\alpha$,0,0)(2$\pi$/a):
a) $\alpha$=1/9, b) $\alpha$=1/3, c)$\alpha$=2/3 (K point). 
The energy loss function is shown 
in the insert. The red (blue) lines represent $MgB_2$ ($B_2$).} 
\end{figure}

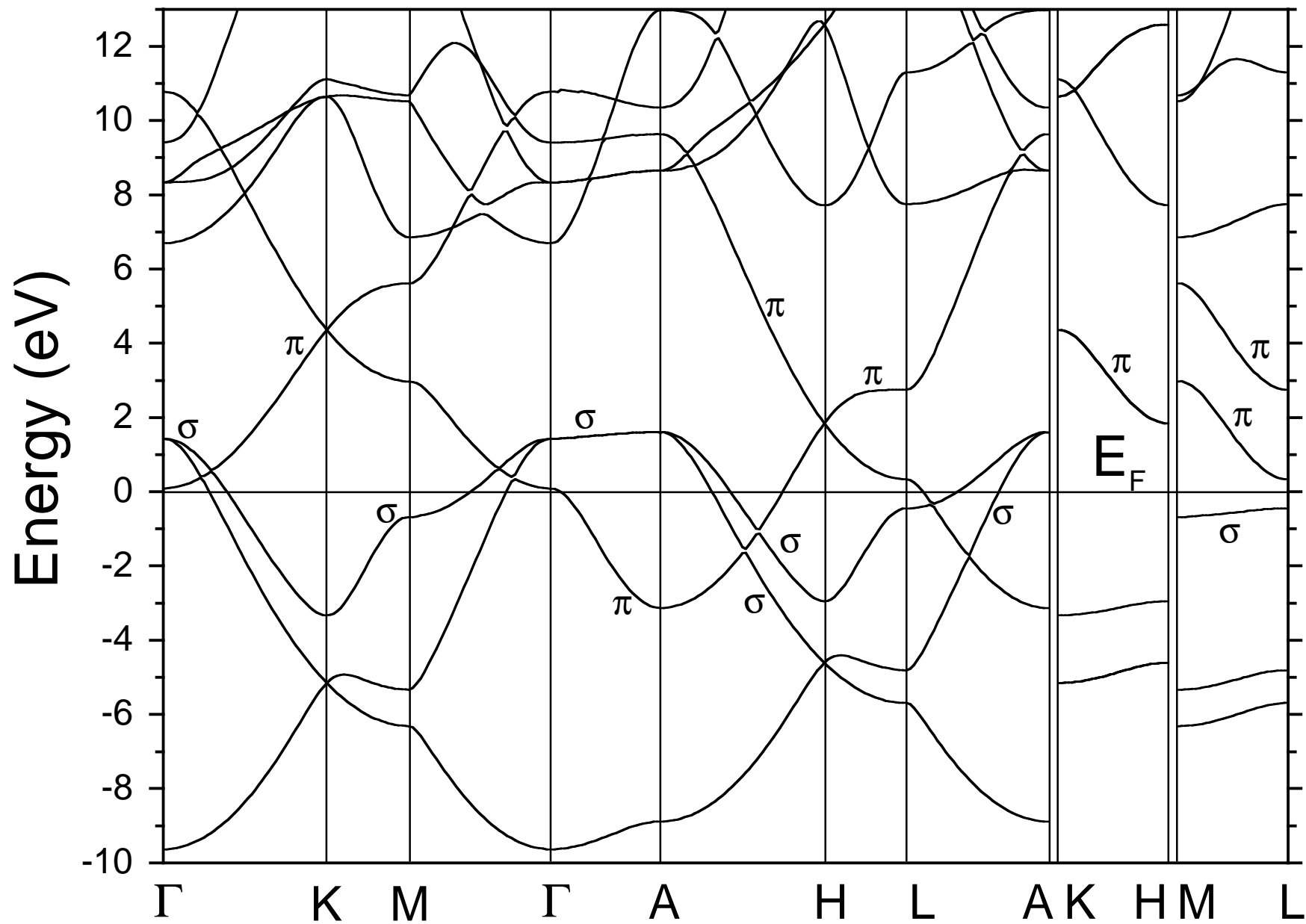
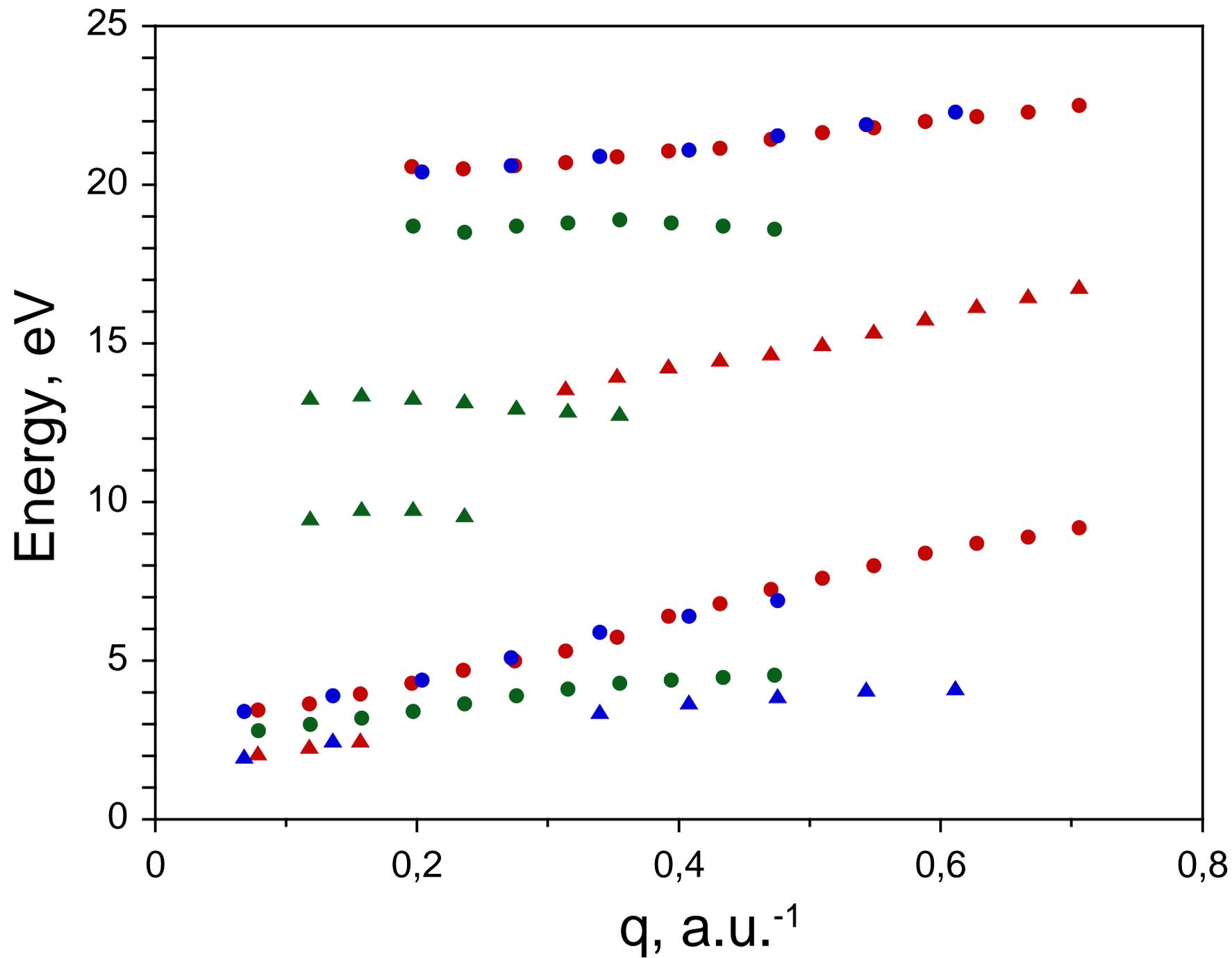
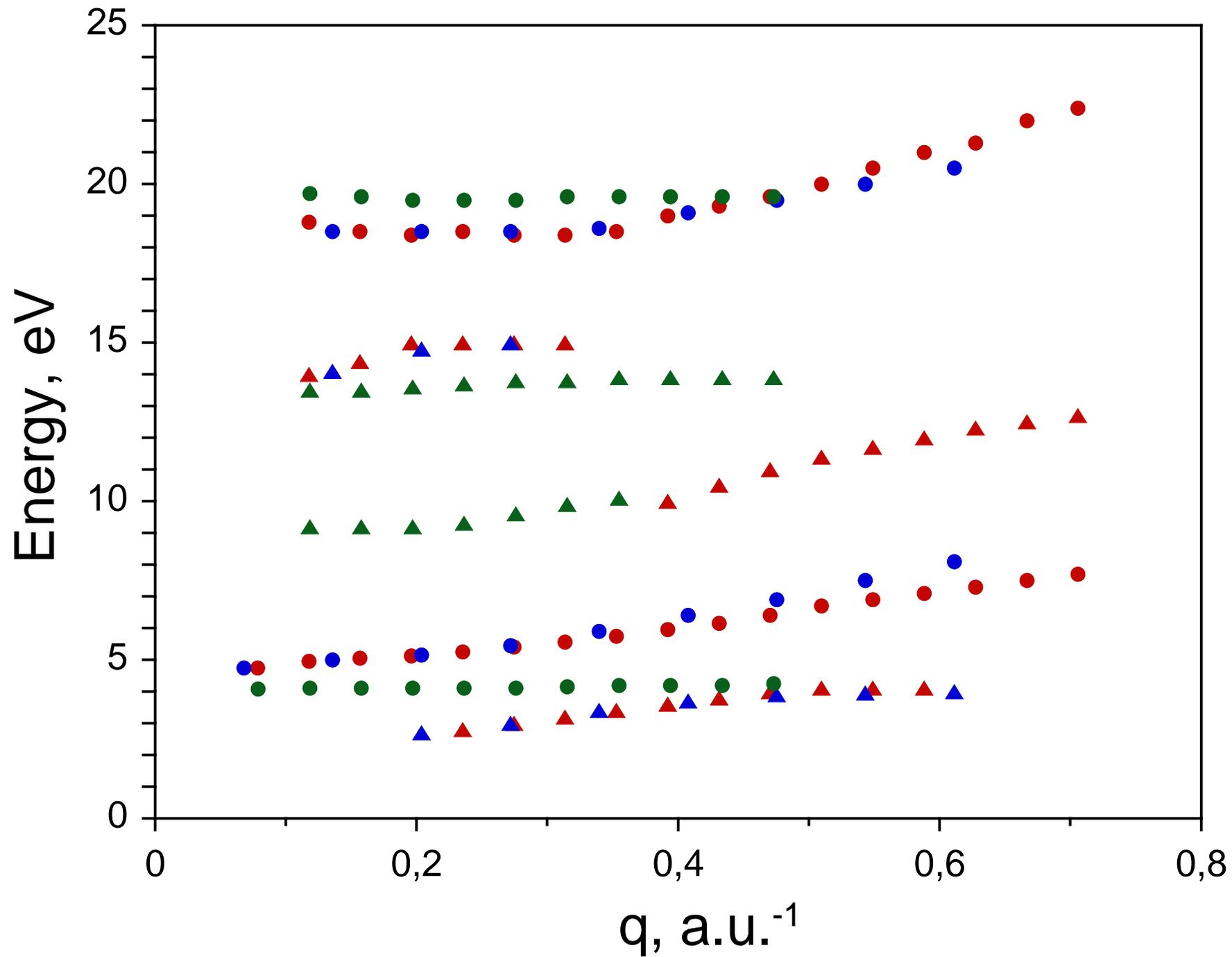
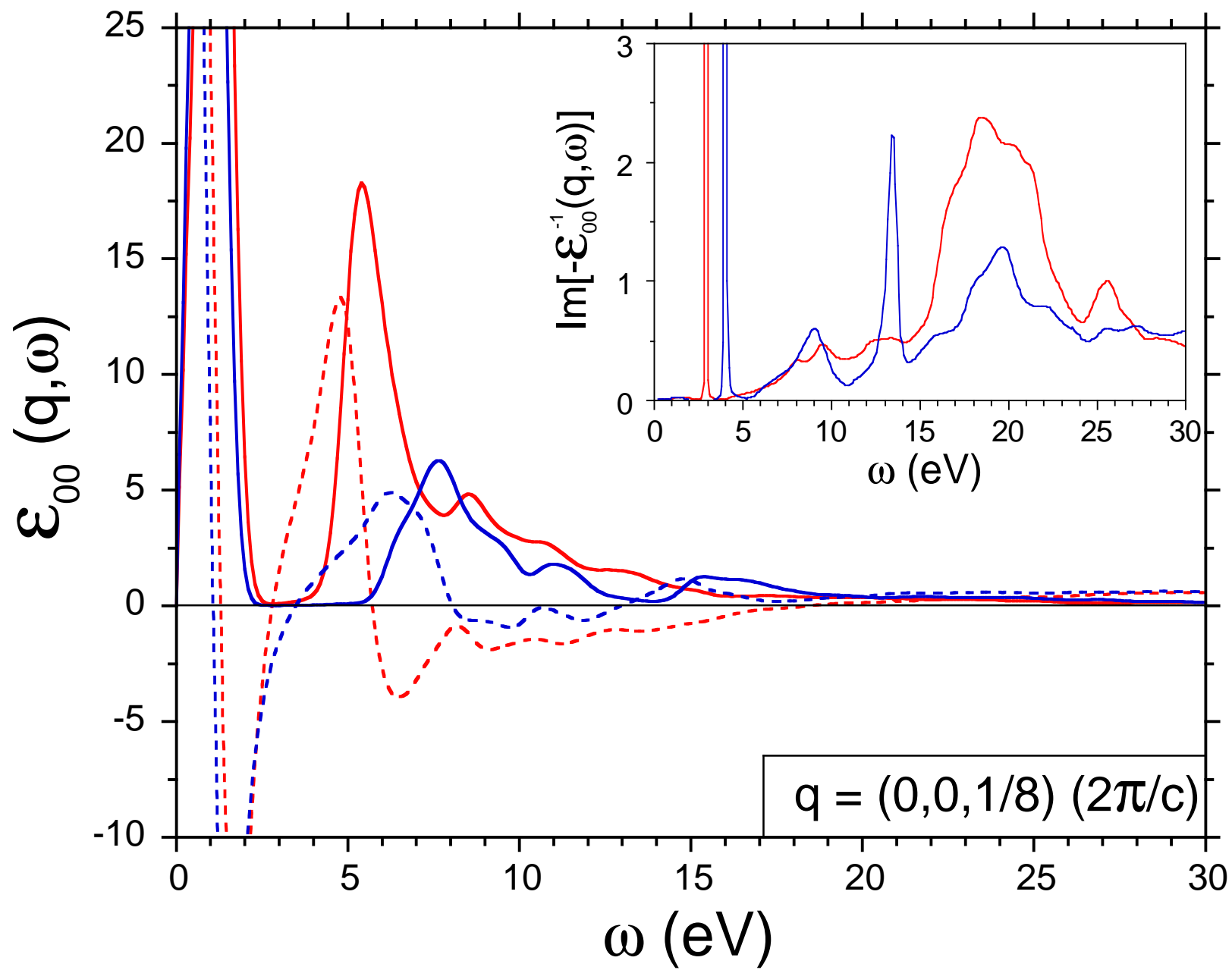
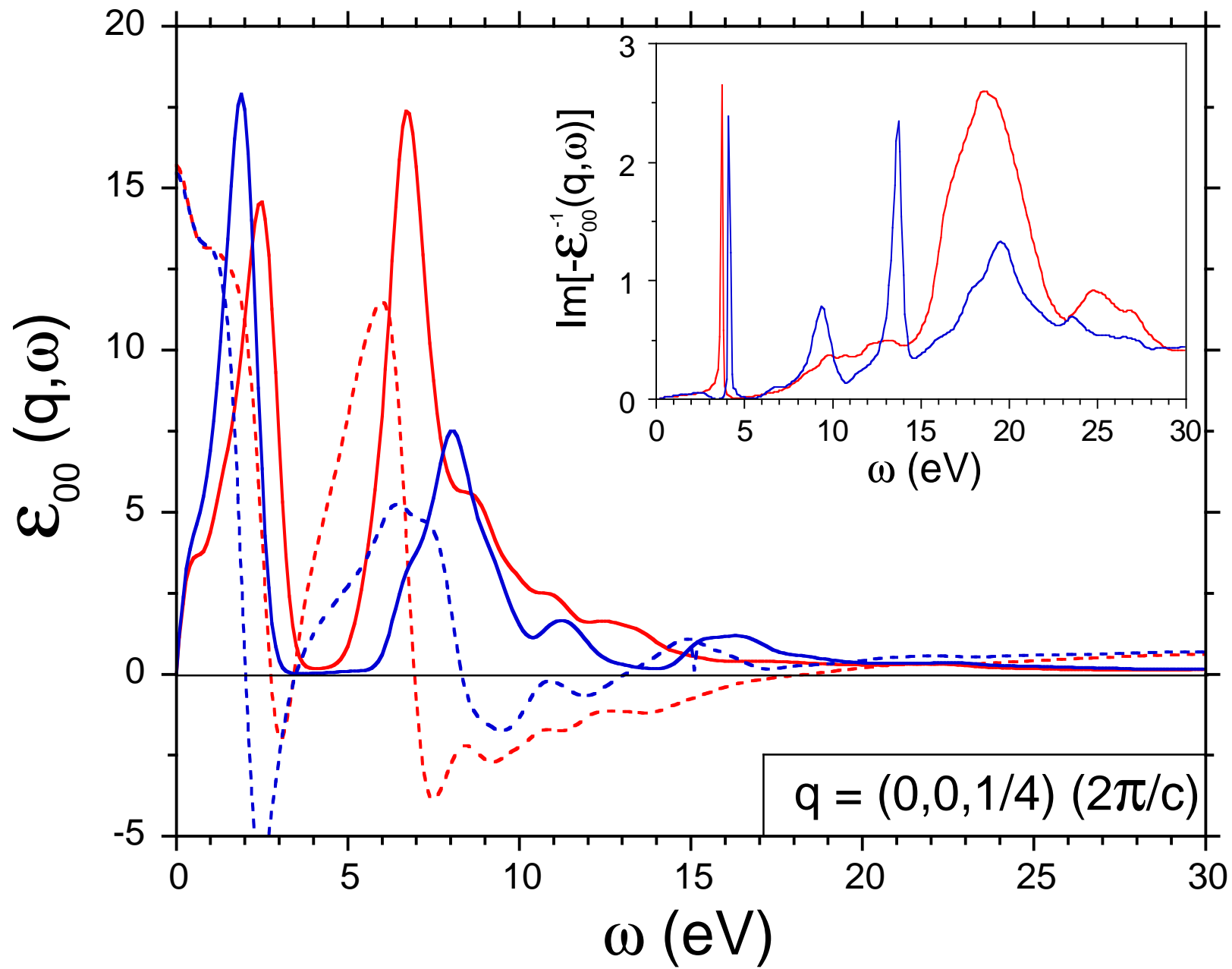
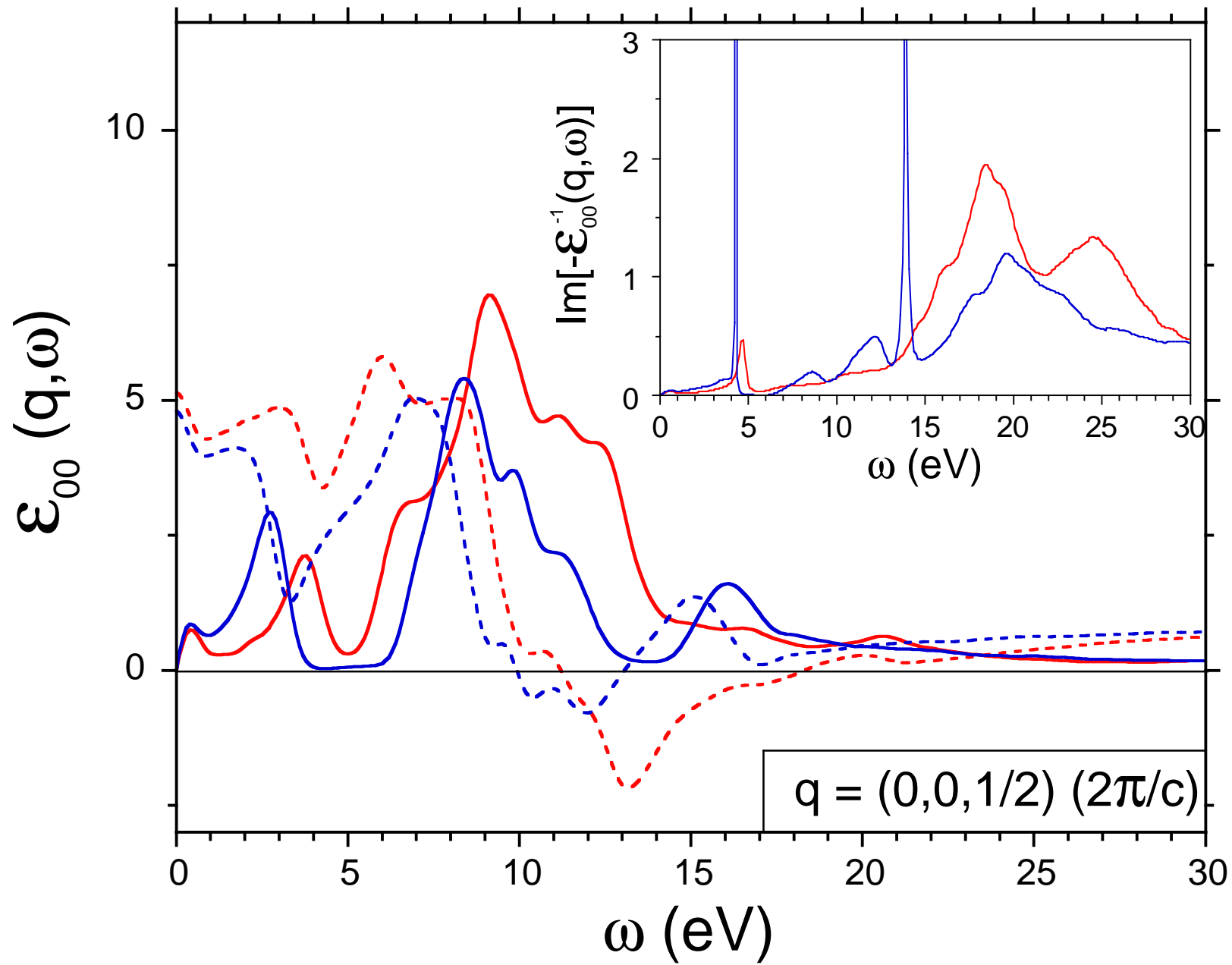
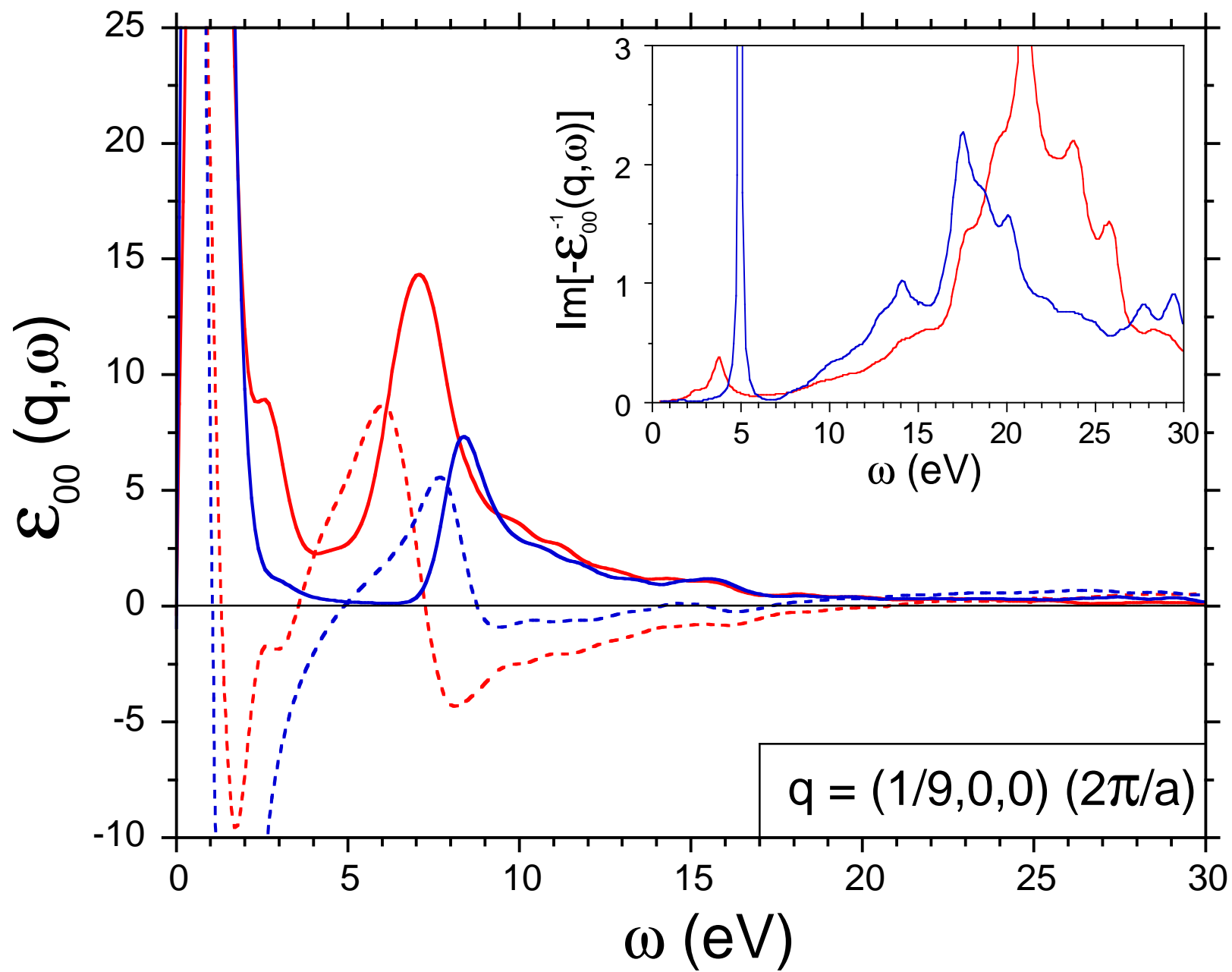
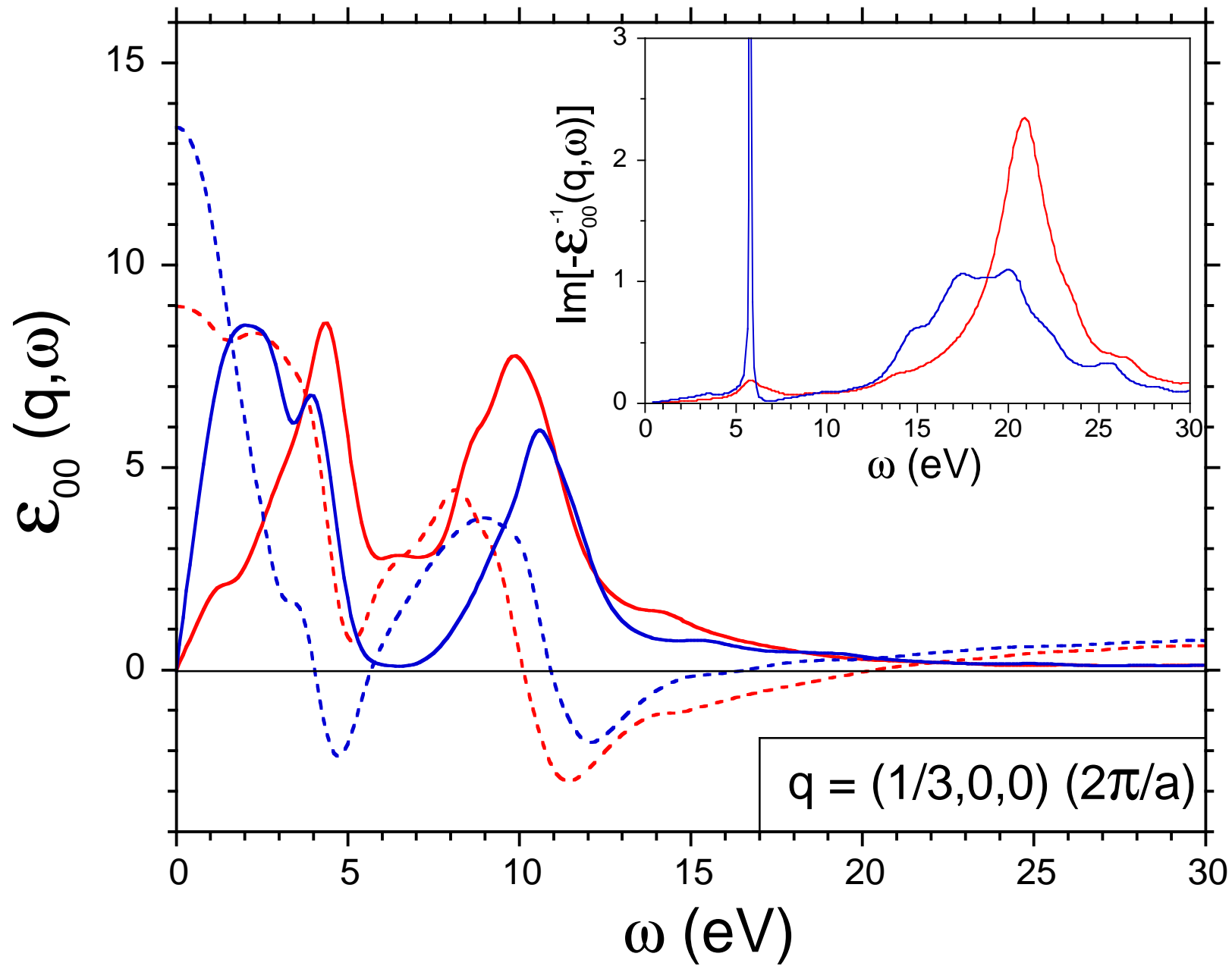
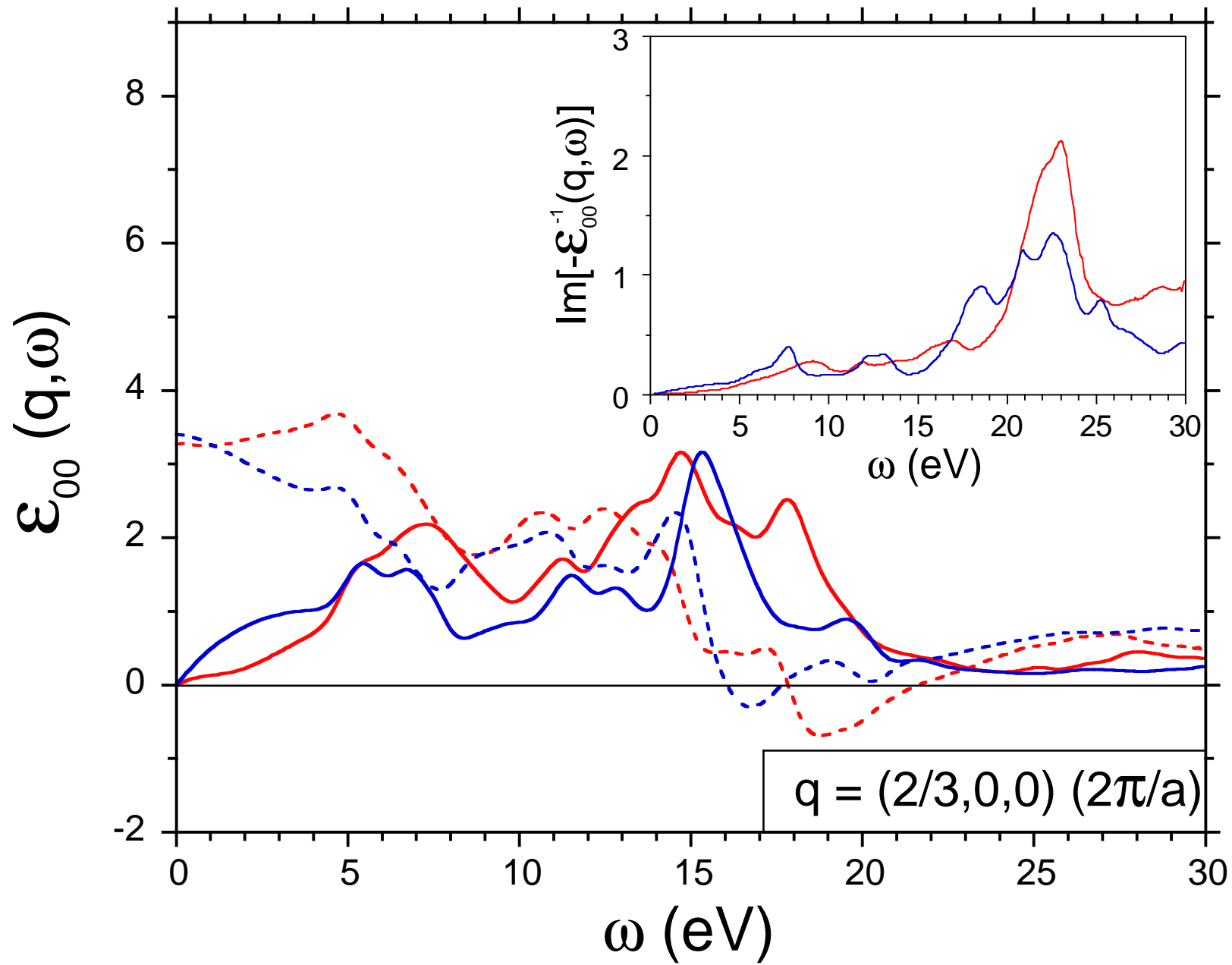
\end{document}